\documentclass[journal]{IEEEtran}

\usepackage{times,amsmath,epsfig}
\usepackage[T1]{fontenc}
\usepackage{graphicx} 
\usepackage{subfigure}
\usepackage{float}
\usepackage{boxedminipage}

\usepackage{cite}

\begin{document}

\title{Frameless ALOHA Protocol for Wireless Networks}

\author{\v Cedomir~Stefanovi\' c,~\IEEEmembership{Member,~IEEE,} Petar~Popovski,~\IEEEmembership{Senior Member,~IEEE,} Dejan Vukobratovi\' c,~\IEEEmembership{Member,~IEEE} 

\thanks{\v Cedomir Stefanovi\' c and Petar Popovski are with the Department of Electronic Systems, Aalborg University,
Aalborg, Denmark (email: \{cs,petarp\}@es.aau.dk).}
\thanks{Dejan Vukobratovi\' c is with the Department of Power, Electronics and Communication Engineering, University of Novi Sad, Novi Sad, Serbia (email: dejanv@uns.ac.rs)}
\thanks{Supported by the Danish Council for Independent Research within the Sapere Aude Research Leader program, Grant No. 11-105159.}
}

\maketitle

\begin{abstract}

We propose a novel distributed random access scheme for wireless networks based on slotted ALOHA, motivated by the analogies between successive interference cancellation and iterative belief-propagation decoding on erasure channels.
The proposed scheme assumes that each user independently accesses the wireless link in each slot with a predefined probability, resulting in a distribution of user transmissions over slots. The operation bears analogy with rateless codes, both in terms of probability distributions as well as to the fact that the ALOHA frame becomes fluid and adapted to the current contention process. 
Our aim is to optimize the slot access probability in order to achieve 
rateless-like distributions, focusing both on the maximization of the resolution probability of user transmissions and the throughput of the scheme. 
\end{abstract}


\section{Introduction}

Slotted ALOHA (SA) \cite{R1975} is a standard random access scheme, in which feedback to the contending terminals is sent after each slot. Framed ALOHA (FA) is a variant in which slots are organized in a frame. Prior to the frame start, each terminal randomly and independently chooses a single slot within the frame to transmit its packet, and it receives feedback at the end of the frame. A typical premise in ALOHA protocols is that the interference among user transmissions is destructive and only slots that contain single transmission are usable. The expected throughput $T$ of the classical SA is $T=G e^{-G}$, where $G$ is the average number of packets sent per slot.
The throughput is maximized for $G = 1$, when $T = 1/e \approx 0.37$.

\begin{figure}[tbp]
	\begin{center}
\includegraphics[width=0.45\columnwidth]{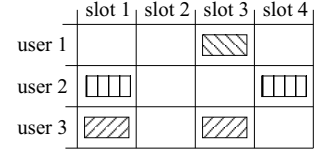}
	\end{center}
\caption{SIC in slotted ALOHA. Packet of user 2 is resolved in slot 4, enabling the removal of its replica in slot 1 and resolution of packet of user 3 in slot 1. In the same way, resolution of packet of user 3 enables the removal of its replica from slot 3, thus resolving packet of user 1.}
	\label{fig:SIC}
	\vspace{-6pt}
\end{figure}

Recently, an important paradigm change was made in \cite{CGH2007}, where the collisions are not considered destructive, as they can be resolved using successive interference cancellation (SIC).
The users repeat their transmissions in multiple slots of the frame and each transmission carries a pointer to the slots where the other replicas take place; this information is used by the SIC algorithm in order to remove replicas of the already resolved packets.
Packet resolution and removal is performed in an iterative manner, as depicted in Fig.~\ref{fig:SIC}.
In this way the throughput is increased; e.g., for the simple scenario when each user performs two repetitions in randomly selected slots of a frame, the throughput is $T \approx 0.55$.

In \cite{L2011} another important upgrade of framed ALOHA has been proposed, based on the fact that the execution of SIC resembles the iterative belief-propagation (BP) decoding on erasure channel, which allows for employing the theory and tools from codes-on-graphs.
The access method of framed ALOHA was further generalized, allowing probabilistic selection of the number of repetitions on a user basis; the related convergence of SIC was analyzed using standard \emph{and-or tree} arguments \cite{LMS1998} and the optimal repetition strategies (in terms of maximizing throughput of the scheme) were obtained.
These strategies resemble the left-irregular LDPC distributions and achieve asymptotic throughput close to 1. 

In this letter we investigate the potential of applying the paradigm of rateless codes \cite{L2002} to the SA framework.
Our objective is to design a random access solution based on analogous principles.
We note that we are not just applying another type of erasure coding (i.e., fixed-rate codes vs. rateless codes), but the rateless analogy brings a major conceptual shift with respect to framed ALOHA: the frame length is not a priori set and becomes fluid, such that new slots are added until sufficiently high fraction of users has been resolved; therefore the name \emph{frameless} ALOHA. This implies that the time instant at which feedback arrives also adapts to the contention process.
The results demonstrate remarkably high throughput values, and the simulation results appear to lead the highest throughput compared to the reported literature.
At the same time, the proposed scheme is rather simple to implement.

\section{Frameless ALOHA protocol}
\label{sec:background}

\subsection{Background: Rateless codes}
\label{sec:rc}

Rateless codes have advantageous erasure-correcting properties as they are universally capacity-achieving, independently of the erasure channel conditions.
They are \emph{rateless} in a sense that the code rate is not a priori set and it depends on the channel conditions: the encoder produces and sends newly encoded symbols until receiving feedback that the message is decoded.
The initial class of rateless codes, LT codes \cite{L2002}, use a simple encoding rule: each encoded symbol is produced by XOR-ing $d$ uniformly and randomly sampled source symbols, where degree $d$ is chosen from a distribution $\Psi(d)$. 

LT codes admit sparse-graph interpretation and iterative BP decoding \cite{L2002}.
In a LT code graph, the source and the encoded symbols represent the left- and the right-side nodes, respectively.
The graph edges reflect the process of combining source symbols into encoded symbols.
An LT code is designed by optimizing the right degree distribution, while the left one asymptotically tends to Poisson distribution due to random uniform source node sampling.
In contrast, by defining the user behavior in random access scheme, one can influence only the left degree distribution, which puts new constraints in the design of LT-like code within the SA framework.

\subsection{Terminology and notation}

\begin{figure}[tbp]
	\begin{center}
\includegraphics[width=0.4\columnwidth]{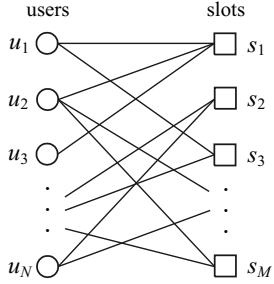}
	\end{center}
\caption{Graph representation of slotted ALOHA with repetitions.}
	\label{fig:graph}
	\vspace{-6pt}
\end{figure}

The relations among users and slots in SA scheme can be represented using a graph (Fig.~\ref{fig:graph}), where the edges connect the users to slots in which their repeated transmissions take place.
We refer to the number of edges incident to user $u$ (slot $s$) as a user degree $|u|$ (slot degree $|s|$).
The user and slot degrees are selected from the degree distributions $\Lambda ( x ) = \sum_j \Lambda_j x ^j$  and $ \Psi ( x ) = \sum_k \Psi_k x ^k$, respectively, where $\Lambda_j = P \big[ |u| = j \big]$ and $\Psi_k = P \big[ |s| = k \big]$.
Further, we introduce edge-perspective degree distributions $\lambda ( x ) = \sum_j \lambda_j x^{j - 1}$ and $\rho ( x ) = \sum_k \rho_k x^{k - 1}$, where $\lambda_j$ ($\rho_k$) is the probability that an edge is incident to a user of degree $j$ (to a slot of degree $k$):
\begin{align}
\label{eq:lambda}
\lambda ( x ) = \Lambda' ( x ) / \Lambda' ( 1 ); \; \rho ( x ) = \Psi' ( x ) / \Psi' ( 1 ).
\end{align}

\begin{figure}[tbp]
	\begin{center}
\includegraphics[width=0.95\columnwidth]{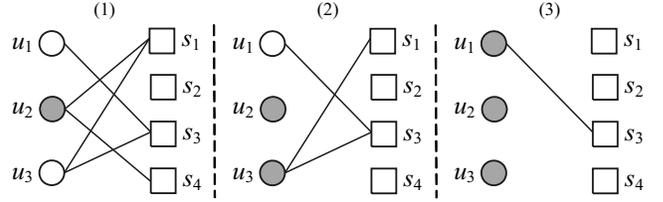}
	\end{center}
\caption{SIC on SA graph. (1) $u_2$ is resolved via the edge $u_2 - s_4$. (2) Edges incident to $u_2$ are erased, lowering the degree of $s_1$ and thus enabling the resolution of $u_3$ via the edge $u_3 - s_1$. (3) Edges incident to $u_3$ are erased, lowering the degree of $s_3$ and enabling the resolution of $u_1$.}
	\label{fig:SIC_on_graph}
	\vspace{-6pt}
\end{figure}

The execution of the SIC algorithm, in terms of SA graph, proceeds as follows.
Initially, the degree-one slots are identified, allowing for resolution of the users (i.e., user packets) connected to these slots and identification of all the other edges (i.e., packet replicas) incident to resolved users.
In the next step, using SIC, these edges are erased from the graph (i.e., replicas are removed), thus lowering the degrees of the incident slots and potentially resulting in new degree-one slots.
These slots enable resolution of yet unresolved users, driving the SIC further.
Fig.~\ref{fig:SIC_on_graph} shows the execution of the SIC algorithm on a graph corresponding to the SA example given in Fig.~\ref{fig:SIC}.

\subsection{System model}
\label{sec:model}

We consider the following setup.
The network consists of $N$ users which contend for the access to the Base Station (BS) with their uplink transmissions ($N$ is assumed known).
The channel is divided into equal-duration slots.
A downlink beacon denotes the start of the contention round, synchronizing the users at two levels:
i) the starting instants of the slots become aligned across users and ii) the numbering sequence of the slots after the beacon becomes a common knowledge across the users.
In each slot every user attempts transmission with a predefined probability, denoted as slot access probability (received via the beacon at the start of the round); this probability is the same for all users in the given slot and, in general, is a function of the slot number.
The round is terminated when the fraction of resolved users reaches a predefined threshold, which is signaled by the next beacon that acknowledges resolved users and initializes the next round.

Hereafter, we assume that the slot access probability $p_m$ that corresponds to slot $s_m$ is:
\begin{equation}
p_m = \frac{G_m}{N},
\end{equation} 
where $G_m$ is the expected degree of slot $s_m$ (i.e., $G_m$ is the expected load of $s_m$). 
The actual degree of $s_m$ is given by the binomial distribution, which can be approximated by the Poisson distribution for the ranges of $N$ and $G_m$ that are of interest in this paper.
It can be shown that:
\begin{equation}
P\big[ |s_m| = k \big] \approx \frac{G_m^k}{k!}e^{-G_m} = \Psi_{mk}.
\end{equation}
The degree distribution of $s_m$ is equal to:
\begin{equation}
\label{eq:Psi_m}
\Psi_m ( x ) = \sum_{k=0}^{\infty} \Psi_{mk} x^k =e^{-G_m ( 1 - x )}.
\end{equation}
The slot degree distribution is LT-like, however, due to the constraints of the access method it can not be optimized directly, as in \cite{L2002}, but only implicitly by controlling the probability $p_m$ (and thus $G_m$), as elaborated in Section \ref{sec:results}.

Suppose that the round totals $M$ slots. 
The average slot degree distribution and the average slot degree are:
\begin{align}
\label{eq:Psi}
\Psi ( x ) = \frac{1}{M} \sum_{m=1}^{M} \Psi_m ( x ),\\
\label{G}
G = \Psi' ( 1 ) = \frac{1}{M} \sum_{m=1}^{M} G_m.
\end{align}

The user degree distribution $\Lambda ( x )$ is also a Poisson one, whose expected value depends on the total number of slots $M$ and average slot degree $G$.
In order to resolve all users, the total number of slots must be equal or greater than the number of users, i.e., $M = ( 1 + \epsilon ) N$, where $\epsilon \geq 0$ is the overhead in number of slots with respect to the ideal case, in which $M = N$.
If we denote the average user degree by $D$ and use the relation $D \cdot N = G \cdot M$, $\Lambda ( x )$ is:
\begin{equation}
\label{eq:Lambda}
\Lambda ( x ) = e^{ -D ( 1 - x ) } = e^{ - ( 1 + \epsilon ) G ( 1 - x )}.
\end{equation}
The main performance parameters of interest are probability of user resolution $P_R$ and throughput $T$, calculated as:
\begin{equation}
\label{eq:throughput}
T = \frac{P_R}{ M / N } = \frac{P_R}{1 + \epsilon}.
\end{equation}
which is a measure of the efficiency of slot usage.
We also note that both $P_R$ and $T$ are functions of $\Psi ( x )$ and $\epsilon$.

\subsection{Analysis}
\label{sec:optimize}

An useful upper bound on $P_R$ can be derived by observing that if a user does not attempt to transmit at all (remains idle during the contention round), he cannot be resolved.
The probability of a user $u_n$ being idle for $M$ consecutive slots is:
\begin{equation}
P\big[ |u_n| = 0 \big] = \Lambda_0 = e^{-(1 + \epsilon ) G} 
\end{equation}
Therefore:
\begin{equation}
\label{eq:bound}
P_R \leq 1 - \Lambda_0 = 1 - e^{-(1 + \epsilon ) G} = P_{UB}
\end{equation}
where $M = ( 1 + \epsilon ) N$.
It could be shown that this bound can be seen as a special instance of the bound derived in \cite{PLC2011b}.
From (\ref{eq:bound}) it can be observed that the greater the average slot degree $G$ and/or overhead $\epsilon$, the lower $P\big[ |u_n| = 0 \big]$.
If $P_R$ is fixed to $P_R = 1 - \delta$, we get:
\begin{equation}
G \geq - \frac{ \ln \delta}{( 1 + \epsilon)}.
\end{equation}

\begin{figure}[tbp]
	\begin{center}
\includegraphics[width=0.9\columnwidth]{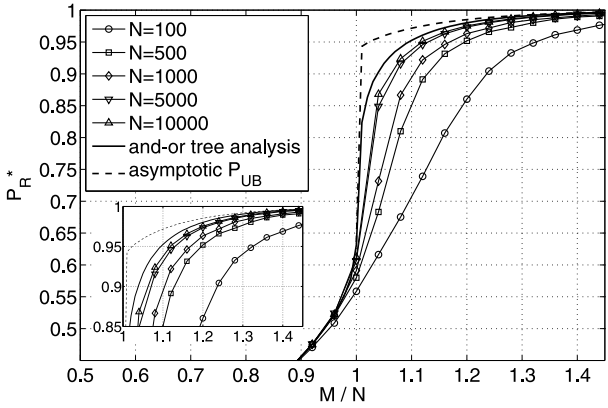}
	\end{center}
\caption{Maximum probability of user resolution $P_R^{*}$ as function of the ratio of number of slots and number of users $M/N$.}
	\label{fig:succ1}
\end{figure}

Consider the slot $s_m$ with the degree distribution $\Psi_m ( x ) = e^{-G_m ( 1 - x )}$.
The probability of slot $s_m$ being idle is then $P \big[ |s_m| = 0 \big] = e^{- G_m } $.
Similarly as before, by increasing the expected degree $G_m$, the probability of $s_m$ being idle is decreased.
However, by increasing $G_m$, the probability that $s_m$ is of low degree decreases as well, which can adversely affect the progress of the SIC.
As elaborated in \cite{L2002}, a substantial fraction of slots is required to be of low degrees in order to drive the execution of SIC.

The and-or tree analysis \cite{LMS1998} can be used to derive the evolution of $P_R$ over the iterations of the SIC algorithm.
Let us denote by $q_i$ ($r_i$) the average probability that an edge in the graph incident to a user (slot) survives the $i$-th iteration.
Using the expressions given in \cite{LMS1998}, as well as Eqs. (\ref{eq:lambda}), (\ref{eq:Psi}) and (\ref{eq:Lambda}), it can be shown that:
\begin{align}
\label{qi}
q_i = & \lambda( r_{i-1} ) = e^{ -(1 + \epsilon ) G ( 1 - r_i )},\\
\label{ri}
r_i = &1 - \rho ( 1- q_i ) = 1 - \frac{1}{GM} \sum_{m=1}^{M} G_m e^{ - G_m q_i}.
\end{align}
The expected probability of user resolution after the $i$-th iteration is $P_R = 1 - q_i$.
By varying $M$ in (\ref{G}), (\ref{qi}) and (\ref{ri}), $P_R$ is obtained for changing number of slots, showing the expected behavior of the proposed approach where the contention round length is not a priori set.

The simplest case is when $G_m$ is constant and equal for all slots $s_m$, i.e., $G_m = \beta$, $1 \leq m \leq M$.
In this case, all the slots share the same degree distribution $\Psi ( x ) = \Psi_m ( x ) = e^{- \beta ( 1 - x )}$, $1 \leq m \leq M$, while the user degree distribution is $\Lambda ( x ) = e^{-( 1 + \epsilon )\beta( 1 - x )}$.
Also, the corresponding expression for $r_i$ is now particularly simple and equal to $r_i = 1 - e^{ - \beta q_i}$.

\section{Results}
\label{sec:results}

\begin{figure}[tbp]
	\begin{center}
\includegraphics[width=0.9\columnwidth]{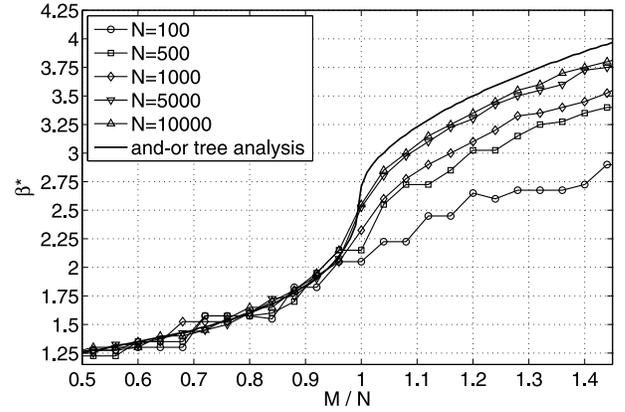}
	\end{center}
\caption{Optimal degree $\beta^{*}$ that yields maximal probability of user resolution $P_R^{*}$, as function of the ratio of number of slots and number of users $M/N$.}
	\label{fig:deg1}
\end{figure}

In this section we present results obtained both using the and-or tree analysis and simulations.
We focus on the simplest case with the constant expected slot degree, as this strategy poses minimal requirements/coordination across the users.
We seek for the optimal degree $\beta^{*}$ that yields the maximum $P_R$ and $T$, denoted by $P_R^{*}$ and $T^{*}$.
Each instance of the optimization is performed for fixed $M/N$ and we vary $M/N$ over the instances.
In the simulation part, we have considered $N=\{100,500,1000,5000,10000\}$ users and all results are obtained averaging over 1000 simulation runs for every $M/N$ of interest.
We have performed only MAC-layer simulations, abstracting the physical layer issues, as justified in \cite{L2011}. 
The SIC was implemented using standard iterative BP decoder for erasure channels \cite{L2002}.

Fig.~\ref{fig:succ1} presents $P_R^{*}$ as function of $M/N$.
The greater the number of users $N$, the closer is $P_R^{*}$ to its asymptotic bound given by the and-or tree analysis.
Also, $P_R^{*}$ displays steep increase after $M/N$ surpasses a threshold value, this threshold being closer to 1 as $N$ grows.
As $M/N$ grows, $P_R^{*}$ saturates and reaches the upper bound given by (\ref{eq:bound}), see the enlarged detail on Fig.~\ref{fig:succ1}. $P_{UB}$ is calculated for $\beta^{*}$ obtained by the and-or tree analysis (i.e., the asymptotic upper bound). 

Fig.~\ref{fig:deg1} shows $\beta^{*}$ that maximize $P_R$.
The optimal value of the expected slot degree grows steadily as $M/N$ increases.
Interestingly, the lower the number of users, the lower $\beta^{*}$ for the given $M/N$ with respect to its asymptotic value.

\begin{figure}[tbp]
	\begin{center}
\includegraphics[width=0.91\columnwidth]{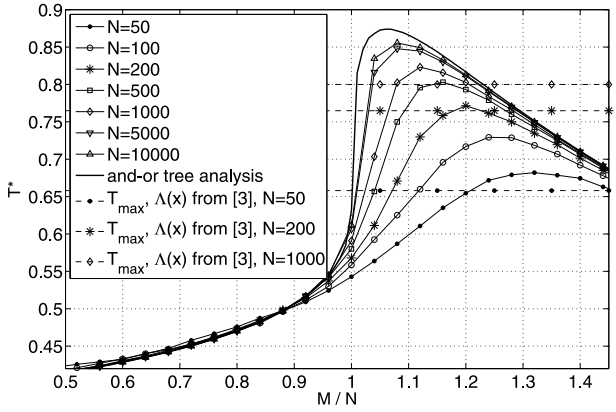}
	\end{center}
\caption{Maximal throughput $T^{*}$ as function of the ratio of number of slots and number of users $M / N$.}
	\label{fig:throughput1}
\end{figure}

Finally, Fig.~\ref{fig:throughput1} shows the throughput $T^{*}$ corresponding to the $P_R^{*}$ given in Fig.~\ref{fig:succ1}, i.e., the maximum throughput obtained for the $\beta^{*}$ for the given $M/N$.
For the sake of comparison, we also included $T^{*}$ obtained for $N=\{50,200\}$.
Initially, as $M/N$ increases, the $T^{*}$ performance follows the $P_R^{*}$ performance, and reaches the maximum value shortly after ratio $M/N$ becomes greater than 1.
As $M/N$ increases further, $T^{*}$ starts to decrease, since the $P_R$ performance saturates while $M/N$ grows, see Eq. (\ref{eq:throughput}).
The maximum asymptotic throughput that can be attained is approximately $0.87$ and it is approached for large number of users $N$.
Nevertheless, the maximum throughput obtained for the realistic number of users is better than the throughput obtained by the more involved methods \cite{L2011}, while the average user degree (i.e., average number of transmissions) is lower.
Particularly, the average user degree obtained for the example user degree distribution $\Lambda(x)=0.5 x^2 + 0.28 x^3 + 0.22 x^8$, given in \cite{L2011}, is equal to $\Lambda'(1)=3.6$, while in the simple scheme presented here the average user degree for $\beta^{*}$ and $M/N$ that maximizes $P_R^{*}$ is always $\Lambda'(1)<3.3$.
At the same time, the maximum throughputs for the presented scheme are consistently better than for the distribution from \cite{L2011}, as showed in Fig.~\ref{fig:throughput1}.

\section{Discussion and practical considerations}
\label{sec:practical}

In practice, the behavior of user $u_n$ over the slots is determined through a function $f(s_m,u_n)$, where $s_m$ denotes the $m$-th slot.
$f(s_m,u_n)$ is a pseudorandom generator tailored to $p_m$ and $N$, which has two outputs - stay silent or transmit, and the slots in which user transmits are completely determined by $u_n$.
We assume that each transmitted packet of $u_n$ carries the identifier of $u_n$, such that upon a successful decoding of this packet, the BS can recover the slots in which $u_n$ has transmitted in the past and will transmit in the future.

The optimized slot access probability $p_m$ depends on the number of users $N$, hence an estimate of $N$ should be available. For a \emph{batch arrival} of terminals, the BS can use a fast estimation algorithm for $N$, as in \cite{KN2006}. For \emph{continuous arrivals} from a user population with $\lambda$ arrived packets per slot, similar to framed ALOHA, $N$ is estimated as $\lambda \tau$, where $\tau$ is the average frame duration, plus the retrying users from the previous frame.
Fig.~\ref{fig:deg1} indicates that $\beta^{*}$ varies slowly with $N$, such that even a rough estimate of $N$ is likely to provide a good throughput, but this observation requires a further study. 

\begin{figure}[tbp]
	\begin{center}
\includegraphics[width=0.87\columnwidth]{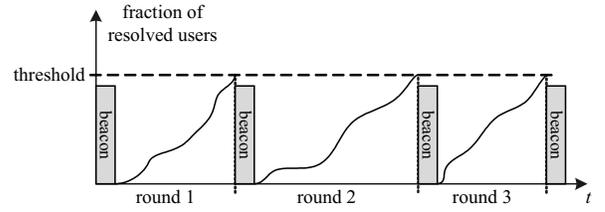}
	\end{center}
\caption{The operation of the proposed scheme over the contention rounds.}
	\label{fig:saw}
	\vspace{-10pt}
\end{figure}

Following Section~\ref{sec:results}, where the varying $M/N$ was implicitly addressed, we investigated the strategy for maximizing the average throughput when the length of the frame (i.e. contention round) is not a priori set, as depicted in Fig.~\ref{fig:saw}.
The contention round lasts until the fraction of resolved users reaches a predetermined threshold, which corresponds to the $P_R$ (Fig.~\ref{fig:succ1}) which maximizes $T$ (Fig.~\ref{fig:throughput1}).
The slot access probability is selected according to $\beta$ that corresponds to the above $P_R$ and $T$.
As an example, for $N=1000$, the threshold is equal to $P_R^{*}\approx0.923$, as this value maximizes $T$, and the slot access probability is $p=\beta^{*}/N=2.9 \cdot 10^{-3}$, as the expected slot degree that maximizes $T$ is $\beta^{*} \approx 2.9$, while the maximum $T$ should be reached for $M\approx 1.1\cdot N = 1100$ slots.
Using this approach, by simulations we obtained that the average length of the contention round is indeed $M\approx 1.1\cdot N = 1100$, with average throughput $T\approx0.83$, as suggested in Fig.~\ref{fig:throughput1}.

The BS is triggered to send the beacon when the fraction of resolved users reaches a predefined threshold. 
The beacon also carries ACK feedback to the resolved users.
In Frequency Division Duplex (FDD) systems, the beacon can be sent through a dedicated downlink channel. In a Time Division Duplex (TDD) system, the beacon could be given precedence using carrier sensing or some other appropriate technique.
These and other practical issues are subject of further work. 


\end{document}